\documentclass[manuscript]{aastex}

\newcommand\Halpha{H$\alpha$}
\newcommand\Hbeta{H$\beta$}

\newcommand\sbunits{$\rm photons~cm^{-2} s^{-1} sr^{-1}$}

\newcommand\cmq{$\rm cm^{-3}$}
\newcommand\cmsq{$\rm cm^{-2}$}

\newcommand\ori{$\theta ^{1}$Ori~C}
\newcommand\oriA{$\theta ^{2}$Ori~A}
\newcommand\emunits{$\rm cm^{-6}~pc$}

\newcounter{ionstage}

\slugcomment{Submitted to the AJ}

\shorttitle{Scattered Nebular Light in the Extended Orion Nebula}
\shortauthors{O'Dell \& Goss}

\begin{document}

\title{Scattered Nebular Light in the Extended Orion Nebula
\footnote{
Based in part on observations obtained at the Cerro Tololo Inter-American Observatory, which is operated by the Association of Universities for Research in Astronomy, Inc., under a Cooperative Agreement with the National Science Foundation.}}


\author{C. R. O'Dell}
\affil{Department of Physics and Astronomy, Vanderbilt University, Box 1807-B, Nashville, TN 37235}
\and
\author{W. M. Goss}
\affil{National Radio Astronomy Observatory, P. O. Box 0, Socorro, NM 87801}
\email{cr.odell@vanderbilt.edu}

\begin{abstract}
We have combined 327.5 MHz radio observations and optical spectroscopy to study conditions in the Extended Orion Nebula. We see a steady progression of characteristics with increasing distance from the dominant photoionizing star \ori. This progression includes a decrease in the F(\Halpha)/F(\Hbeta) ratio, an increase in the relative strength of scattered stellar continuum, decrease in electron density determined from the [S~II] doublet, and increase in the ratio of Emission Measures derived from the \Hbeta\  line and the 327.5 MHz radio continuum. We conclude that beyond about 5' south of \ori\ that scattered light from the much brighter central Huygens region of the nebula significantly contaminates local emission. This strengthens earlier arguments that wavelength and model dependent scattering of emission line radiation imposes a fundamental limit on our ability to determine the physical conditions and abundances in this and arguably other similar Galactic Nebulae. The implications for the study of extra-galactic H~II regions are even more severe. We confirm the result of an earlier study that at least the eastern boundary of the Extended Orion Nebula is dominated by scattered light from the Huygens region.

\end{abstract}


\section{Introduction}
The vast majority of observations of the Orion Nebula (NGC~1976) have been made of the bright central region originally depicted by Huygens \citep{gin} . This region around the Trapezium stars and extending to the Bright Bar feature to the SE  is more than two orders of magnitude higher surface brightness than parts of the region designated \citep{gud} as the Extended Orion Nebula (EON). The EON is an elliptical structure of 30\arcmin\ x 27\arcmin\ ($\sim$3.9 pc x 3.5 pc) oriented towards a position angle of 35\arcdeg, with the Huygens region ($\sim$5\arcmin, 0.7 pc) lying in the NE corner. The EON is bounded by an irregular but usually well defined edge that is somewhat obscured in the NE portion by extinction from the foreground Veil of primarily neutral material \citet{vdw89,abel}. The most visible component of the Veil is the Dark Bay lying to the east of the Trapezium.  

The EON has been imaged many times (e.g. Spitzer Space Telescope press release ssc2006-21a, Robert Gendler's site http://www.robgendlerastropics.com/Nebulas.html) at progressively higher spatial resolution (the highest resolution being with the Hubble Space Telescope \citep{wjh}), but never with clear spectral resolution of the group of strongest lines in the red ([N~II] 6548 \AA, \Halpha\ 6563 \AA, [N~II] 6583 \AA). This limitation has made it difficult to quantitatively determine the physical conditions in the EON, an important shortcoming because it is now known that the EON hosts at least two extended regions of million degree gas \citep{gud}. The arguably best attempt to interpret wide bandpass observations is the study of \cite{sgm}, where photographic images were compared with Very Large Array (VLA) images at long wavelengths. The present paper reports on work that can be considered an extension of that study, with the important difference that we use calibrated spectra to determine the optical region characteristics of the nebula. 

In this paper we report on a portion of a large data set of spectroscopy of the EON. The full set of data will be published after completion of spectroscopic mapping of the entire EON.  The centers of the spectroscopic samples used here extend out to 14.0\arcmin\ south of the Trapezium and 17.6\arcmin\ to the west.  There have been other spectroscopic studies that went beyond the Huygens region, the most notable being \citet{jps73} (out to $\sim$7.2\arcmin\ from the Trapezium to the northwest),  \citet{pp77} (out to $\sim$3.7\arcmin\ from the Trapezium to the south),  \citet{oh65} who combined photographic spectroscopy and narrow-band filter photometry and included one point 19.5\arcmin\ to the SW, and the farthest sample of the spectrophotometric of the Huygens region by \citet{b91} was 4.9\arcmin\ from the brightest star in the Trapezium \ori .

\section{Observations}\label{obs}
We draw on two sets of observational data in this study. The first are a new set of spectrophotometric measurements with a visual wavelength range, long-slit spectrograph in selected regions extending out into the EON. The second set are VLA observations used in a previous study \citet{sgm} at 92 cm (327.5 MHz).

\subsection{New Spectrophotometric Observations}
\subsubsection{The Observations}
New spectroscopic observations were made at the Cerro Tololo Interamerican
Observatory 1.5 m telescope operated in cooperation with the SMARTS consortium.
The instrument was the Boller and Chivens spectrograph. Observations were made in 2008
November 22 (positions P1363 and P1605) and 2008 November 23 (positions JW75 and JW887) with Grating G58 using the
 Loral 1K CCD detector and on 2008 November 25
(positions S240 and S360), and 2009 January 16 (positions S420, S480, S600, S720, and S840) with the G09 grating.
In all cases a GG395 glass filter was used to prevent second order
flux from contaminating the first order images that were targeted.
On all nights one pixel projected 1.3\arcsec\ along
the 429\arcsec\ long slit, while
the slit width was 2.6\arcsec\ during the G58 nights of observations
and 3.4\arcsec\ on the G09 nights. The measured full width at half maximum
intensity of the nebular lines were 6 \AA\ on nights with G58, and
7 \AA\ on nights with G09.
The CTIO spectrophometric standard star Feige 15 was observed on each night.
In all cases the sky spectrum was measured periodically at locations well
removed from the EON and subtracted from the nebular spectra.
%extended Orion Nebula (EON, a term first
%employed by \citet{2008Sci...319..309G} to designate the elliptical, low surface
 %portions of the Orion Nebula that extend to the southwest).

The positions of the slit settings
for each of these nights are shown in Figure \ref{fig:slits}.  Where a star other than \ori\ (5:35:16.4 -5:23:23.0 2000) was used for reference, JW prefixes indicates a star in the \citet{jw88} catalog and P indicates a star in the \citet{par54} catalog. JW75 was located 54.6\arcsec\ east of the center of the JW75 spectrum, the east end of the JW 887
spectrum was at 5:35:52.52 -5:32:23.0 (2000), star P1353 was 5.3\arcsec\ east of the center of the P1353 spectrum, star P1605 was 9.1\arcsec\ north of the center of the P1605 spectrum, \ori\ was 1.4\arcsec\ west of the center of the 
S240 and S360 spectra (which are displaced 240\arcsec\ and 360\arcsec\ south of \ori) and \ori\ was 42.4\arcsec\  east of the center of the S420, S480, S600, S720, and S840 spectra, which were similarly
displaced south of \ori. The slit center positions are given in Table 1.

In order to facilitate sky subtraction and cosmic ray cleaning multiple expsoures were made. The exposure times were for each spectrum (JW75 6x900 s, JW887 6x900 s, P1353 4x1800 s, P1605 4x1800 s, S240 2x300 s, S360 2x300 s, S420 through S840 2x600 s).

\subsubsection{Data Reduction}

Data reduction was done using standard IRAF procedures\footnote{IRAF is distributed by the National Optical Astronomy Observatories, which is operated by the Association of Universities for Research in Astronomy, Inc.\ under cooperative agreement with the National Science foundation.}. The results of
these steps were calibrated spectra expressed in ergs \cmsq\ s$^{-1}$ pixel$^{-1}$, which were then converted to surface brightness units and averaged over the entire length of the entrance slit. Only the 
\Halpha\ 6563 \AA, \Hbeta\ 4861 \AA, [N~II] 6583 \AA,  [N~II] 6548 \AA, [S~II] 6716 \AA, and [S~II] 6731 \AA\ lines were measured using task "splot", which required de-blending the lines near \Halpha\ and the [S~II] doublet using task "deblend". The spectra are rich in many other high signal to noise ratio emission 
lines and a more complete compilation of the results will be published later.  In addition to the emission lines, we have also determined the equivalent width of the underlying continuum at the wavelength of \Hbeta. The results from two spectra obtained during the same sequence of observations were used to compare the derived surface brightness in the \Hbeta\ line with those obtained from the spectrophotometric study of \citet{b91} and the calibration of the Hubble Space Telescope WFPC2 emission line filters \citet{od99}, which use the \citet{b91} results as a standard, and it was determined that our results were lower. We have, therefore, applied a normalizing  factor of 1.21 to our surface brightness values and these are given in Table 1, in addition to a summary of results for other observed values. In all cases the entries of Table 1 are averages over the entire slit length and the distance from \ori\ is to the center of the slit sample.

The normalized surface brightness in the \Hbeta\ line  S$_{H\beta}$ (photons  \cmsq\ s$^{-1}$ steradian$^{-1}$) was converted to Emission Measures (EM)  by the relation
EM= (4$\pi /\alpha^{eff} _{H\beta}$) S$_{H\beta}$ .  Using an interpolated value of 3.63x10$^{-14}$ cm$^{3}$~s$^{-1}$ from Table 4.2 of  \citet{agn3} for 9000 K gives EM= 1.122x10$^{-4}$~S$_{H\beta}$   \emunits. These derived EM values are not corrected for any interstellar extinction. The results are presented in Table 2 as are the electron densities derived from the observed [S~II] nebular doublet
emission line ratios given in Table 1 and using the IRAF/STSDAS ''temden" package \citep{sd94}. We have again used an electron temperature of 9000 K, which is characteristic of these regions, as determined by a comparison of auroral and nebular transitions of [O~III] and [N~II] and this temperature is similar to that of the inner, Huygens region of the nebula
\citep{opp}.

We have also derived a pixel by pixel determination of the profile of the EM along a combination of the JW887 and the S480 slits, which are within a few arcsecs of one another in declination and overlapped by 33\arcsec. The composite profile was 825\arcsec\ long, extending west from 5:35:54.52 -5:31:23.0 (2000). This profile was then convolved with a gaussian of full width at half maximum (FWHM) intensity of 67\arcsec\ in order to match the resolution along the east-west axis of the complementary 327.5 MHz images.

\subsection{The 327.5 MHz VLA Observations of Orion A}

The radio observations used in this study are those presented by  \citet{sgm}; the resolution is $78.6\arcsec \times 65.0\arcsec$ at position angle $25^{\arcdeg}$. The rms noise is 7.6 {mJy~beam$^{-1}$. The peak in the VLA image at  327.5~MHz is 3.23 Jy~beam$^{-1}$. Details of the imaging process are
  given by \citet {sgm}. The contour image is shown in Figure 2. This wide field image includes NGC 1976 (M~42) and NGC 1982 (M~43) as well as NGC 1973-75-77 located about half a degree north of M42.

The brightness temperature of the 327.5~MHz radiation has been converted
to EM based on the assumption that the emission is optically thin. The standard equations were used \citep{mez67} with an assumed electron temperature of 9000 K. The assumption of small optical depth is justfied since the brightness temperatures at 327.5~MHz range from $\sim$20~K to $\sim$400
K along the slice. Each sample represents the average EM over the VLA beam of FWHM $ \sim$ 71 \arcsec. The average EM at 327.5~MHz given in Table 3 are the average  EM over a range $\pm$ 215\arcsec\ in right ascencion or declination  centered at the positions of the relevant optical \Hbeta\ samples. The EM slice from the VLA data shown in Figure 6 was determined by sampling the FWHM $\sim$ 71\arcsec\ beam at about two pixels per resolution element. The zero point of the slice was determined from the absolute J2000 coordinates of the \Hbeta\ data. 

\section{Discussion}
\subsection{The Model for the Orion Nebula}
The widely accepted model for the Huygens region of the Orion Nebula is that of a thin blister of ionized gas on the observer's side of a giant molecular cloud, a model that has been continuously refined in detail \citep{od01} and has even allowed the construction of a 3-D model \citep{ow94}, with the dominant ionizing star \ori\ lying about 0.2 -- 0.3 pc in front of the Main Ionization Front (MIF, the surface at which the nebula has become optically thick to Lyman continuum (LyC) radiation). The MIF is essentially concave, with a nearly perpendicular portion to the southeast from the Trapezium that causes the nearly linear Bright Bar feature.  Most recently it has been argued that the bright Orion-S center that lies  to the southwest of the Trapezium is actually a detached portion of optically thick gas lying in front of the surface blister of ionized gas \citep{od09}. In front (the observer's side) of the low density region containing most of the stars of the Orion Nebula Cluster lies the foreground Veil \citep{vdw89,abel}. The Veil is mapped through 21 cm absorption lines seen against the free-free radio emission from the Huygens region and the extinction it produces, with a good correlation between the two \citep{od92}. The Veil is thought to lie about one pc \citep{abel} in the observer's direction from \ori, is optically thick to LyC radiation, and has the greatest column density to the northeast around the Dark Bay feature. It is not traced farther to the NE because of a lack of background emission necessary for determining the atomic hydrogen column density.  

The structure of the EON is basically unknown, although addressed in \citet{sgm}, a paper supplanted by the present study.  The structure appears to be an elliptical cavity, bounded around the oval perimeter, and has recently been determined to contain two large masses of low density X-ray emitting hot gas \citep{gud}.  The amount of material on the observer's side of the EON is uncertain, but the far side is almost certainly a continuation of the MIF and is a blister of ionized gas.  Beyond the MIF in the EON and the Huygens region must be a higher density region of dust and gas known as the photon dominated region (PDR), visible only through infrared and radio emission, but the dust component just beyond the MIF is expected to be optically thick to visual wavelength radiation (the nebula becomes optically thick to hydrogen atom absorption of  LyC radiation before becoming optically thick to scattering by dust)  and this dust can back-scatter towards the observer both incoming star and nebular light \citep{od01}. The scale of this nearest H~II region is about 0.13 pc/arcmin, using a distance of 440 pc \citep{oh08} as derived from multiple studies of the region. 

\subsection{Comparison of EM values derived from \Hbeta\ and 327.5 MHz emission} 

Comparison of the EM values derived from the radio continuum and the \Hbeta\ emission can illuminate the origin of the emission seen in various parts of the EON.  In the absence of other processes, the EM derived from both \Hbeta\ and the radio continuum should be the same. If they are not, it means that additional mechanisms are modifying the radiation that is observed. Interstellar extinction would diminish the surface brightness in \Hbeta\ but leave the 327.5 MHz radiation unaffected. This means that the EM derived from \Hbeta\  would be artificially low and one would expect that the EM ratio  derived from the two wavelengths, EM(\Hbeta)/EM(327.5 MHz) would be less than unity. The expectation would be that all values of the ratio would be less than or equal to unity, whereas examination of the last column of Table 3 shows that only the JW75 sample is less than unity and many
of the values are much larger. 

The EM ratio results are also plotted in Figure 3, where it is seen that there is a systematic change with distance from \ori, with the ratio increasing with distance. All of the samples taken to the south show this pattern, with only the two points on a westward track  (JW75 and P1353) not showing the pattern. Studies of the extinction in the Huygens region have demonstrated that almost all of the extinction arises within the Veil \citep{od92}, rather than the intervening interstellar gas, and that it diminishes as the line of sight moves away from the Dark Bay \citep{oys}. Therefore the trend in the ratio is completely contrary to that expected from variations in the extinction.

It is necessary to consider the effects of the two very different resolutions of the two data samples. The 327.5 MHz data has FWHM=71\arcsec\ while that of the \Hbeta data is a few arcseconds. This difference is 
probably not important in the comparison of the radio and optical results because of the large size of our sample apertures (length 429\arcsec). In addition, if we have not normalized our observations to the surface brightness values of \citet{b91} our EM ratios would all be reduced by a factor of 0.83 and the onset of the excess EM ratio would be shifted outward slightly.

\subsection{Scattering from PDR dust as the explanation of the anomalous EM ratios}

Several studies have already established that emission line radiation arising from near the MIF in the Huygens region can be scattered by dust in the nearby PDR. This effect is indicated from the polarization of the radiation \citep{ll87} and the fact that difference of velocity of the emitting and scattering region produces a red shoulder on the line profiles when observed at high spectral resolution \citep{od92,hen94,hen98}. About 20\%\ of the line radiation from within the Huygens region is scattered light \citep{od01}. The determination of any wavelength dependence of this scattering has not been made but could be made by high resolution study at multiple wavelengths of a series of intrinsically narrow lines arising from the same distance from the PDR. The wavelength dependence of the back-scattering can only be established theoretically if there is a very exact knowledge of the absorption and scattering properties of the particles, their size distribution, and their location; however, it can be expected that the scattering is relatively more efficient at shorter wavelengths in the same manner that there is an observed reddening in the general interstellar extinction.

There is a PDR behind the extension of the MIF as the line of sight is extended into the EON since the background molecular cloud is much larger than the EON. The existence of such a PDR means that the same mechanism that scatters emission lines in the Huygens region should also be operating there. However, the scattered light in the Huygens region is almost all local, that is, there is little separation of the emitting gas and the scattering dust, whereas in the outer regions one must consider scattering of radiation that does not arise locally. 

The brightest parts of the Huygens region have S$\rm _{H\beta} =$1.4x10$^{10}$ \sbunits. 
The highest surface brightness sample in our study is S240 (S$\rm _{H\beta} =$4.69x10$^{8}$ \sbunits) 
and the faintest is S840 (S$\rm _{H\beta} =$5.12x10$^{7}$ \sbunits), a range from 3\%\ to 0.4\%\ of the brightest parts of the Huygens region. If the scattering efficiency of the Huygens region ($\sim$20\%) applies in the EON, there is expected to be a significant portion of the radiation observed there will actually be scattered light, rather than emission arising locally. 

An observed EM ratio of greater than unity would then be explained as S$\rm _{H\beta}$ being ''contaminated'' by scattered \Hbeta\ arising from the Huygens region, with the amount of contamination being indicated by the ratio. If the EM ratio is two, the there are equal amounts of locally emitted and scattered radiation. If there is local extinction, then the contamination indicated by the EM ratio will be an underestimate.

The rise in the EM ratio with increasing distance for the samples to the south of the Huygens region indicate that the local intrinsic surface brightness decreases even more rapidly. The small amounts of scattered light in the samples to the west indicate that those observations are less affected by scattering. The difference in the south and west samples probably also indicates different conditions. Both west samples (particularly JW75) lie approximately on a line from the brightest parts of the Huygens region and the Orion-S cloud which is thought \citep{od09} to lie in front of the MIF emitting layer. This trend means that they could be partially in a shadow of the optically thick Orion-S cloud.  

\subsection{Evidence for scattering of star light}

The continuum in the nebular spectra of the Huygens region is dominated by scattered star light. As noted in \S\ 2.1.2, the relative strength of the \Hbeta\ emission line and the underlying continuum is best expressed by the  equivalent width (EW), the wavelength interval of the continuum required to give as much flux as the \Hbeta\ emission line. A smaller value of EW means a stronger continuum relative to \Hbeta. In the case of the high density Huygens region the EW from atomic processes should be about 1800 \AA\ \citep{od01}, a value that should decrease only slightly in lower density regions where the contribution from two photon decays of hydrogen become a more important contributor to the continuum. In the east-west slice study of  \citet{b91}, the EW slowly changes with increasing distance from \ori, as shown in Figure 4. The current data samples are from farther out and are also shown in Figure 4. The photoelectric filter photometry samples of \citet{oh65} were within the Huygens region, with an outer sample at 19.5\arcmin\ to the south on the rim of the EON. The variation of EW with distance from these three studies indicates that scattered star light becomes relatively more important with distance. The exception to the general pattern is again with samples JW 75 and P1353. Since they lie in a region that may be shadows by the Orion-S cloud, it is not surprising that their EW values are larger.  

The most elementary view would be that the EW should be constant with distance, since at this level of modeling the emission from \Hbeta\ would scale with the ionizing flux from \ori\ and the scattered light would scale with the continuum flux at 4861 \AA. Part of the cause of the variation in EW may be the fact that the LyC and optical flux do not diminish at the same rate with increasing distance from \ori. This is because the ratio of neutral hydrogen atoms per grain will increase with increasing distance as the fraction of ionized hydrogen decreases with distance. The other possibility is that the continuum is a composite of scattered light from all stars. The next three brightest stars in the Trapezium contribute 51\%\  the visual luminosity of \ori\, with unknown distances along the line of sight.  \oriA\  is 4\%\ brighter visual brightness than \ori, but again we don't know its location along the line of sight. This uncertainty in position leaves open the possibility that the samples extending to the south may contain an important amount of \oriA\  scattered light. Attempts to find variations in the relative strengths of the stellar absorption lines in the scattered light continuums were not successful because of inadequate signal to noise ratio.

Clearly scattered star light is very important in the Orion Nebula, this scattering almost certainly occurs in the dust in the PDR, just beyond the ionized gas of the nebula. We should not be surprised to also see evidence for scattered emission lines. 

\subsection{Interpretation of F(\Halpha)/F(\Hbeta) ratios as wavelength dependent  scattering}

Comparison of the observed F(\Halpha)/F(\Hbeta) ratios shown in Table 1 indicate that almost all of the observed ratios are below the value of 2.89 \citep{agn3} expected for an electron temperature of 9000 K. 
We noted in \S\  3.4 that it is reasonable to expect that there will be a wavelength dependence of the scattering in the PDR particles. This is supported by the photoelectric filter study of the continuum in the Orion Nebula.  Table 3 of \citet{oh65} shows that the continuum (which is dominated by scattered light) grows steadily bluer as the line of sight moves away from the Trapezium. 

There is a general pattern of decrease in the F(\Halpha)/F(\Hbeta) with increasing distance from \ori\ in multiple studies, as summarized in Figure 5.  The most anomalous with respect to the general decrease is for P1353, which lies a great distance to the west, where conditions must be different from those in the southern samples.  Intermediate distance points from \citet{oh65} have not been shown because these data points did not have an independent determination of contamination of the filter signal from the [N~II] doublet.  

The rapid change at smaller distances primarily reflects decreasing optical depth in the foreground Veil as the line of sight moves away from the highest opacity Dark Bay. However, the decrease below a value of 2.89 must indicate the increasing importance of wavelength dependent scattering. This scattering probably also plays a role even before the value of 2.89 is crossed  at a distance $\sim$5\arcmin.  Figure 4 indicates that scattered light from the Trapezium has already become more important by that distance and Table 3 shows that the S240 sample  has an EM ratio more than unity.

\subsection{EM ratio change along an east-west sample}

Because of the near coincidence in declination of samples S480 and JW887, we have been able to make an 825\arcsec\  long profile of the surface brightness in \Hbeta\ and the EM.  This sample was convolved to the same resolution as the 327.5 MHz  profile described in \S\  2.2 and the results for both the optical and radio derived EM values are shown in Figure 6, with the ratio of the values presented in Figure 7. In these figures we see that the EM ratio is about 1.5 for the region west of the North-South Rim, but rises precipitously to the east of that feature. The EM ratio (Figure 7) rises rapidly as determined by the low resolution of the radio data (FWHM=67\arcsec), and is consistent with most of the radiation to the
east of the North-South Rim arising from scattered light that has originated in other regions.

This interpretation is strengthened by a detailed analysis of the JW887 sample. For a sub-sample of the eastmost 168\arcsec, F(\Halpha)/F(\Hbeta)= 3.04 , F(6716 \AA)/F(6731 \AA)=1.14, EW=160 \AA, and the electron density from the [S~II] doublet is 320 \cmq. The corresponding values for the remainder of the JW887 sample are F(\Halpha)/F(\Hbeta)=2.77, F(6716 \AA)/F(6731 \AA)=1.29, EW=330 \AA, and the [S~II]  density is 130 \cmq.  We see a large increase in the strength of the scattered starlight continuum and a Balmer line ratio in the east sample that is similar to that of a sample 
taken at half the distance from \ori.  Since much of the Balmer line signal is from scattered light, the value of F(\Halpha)/F(\Hbeta)= 3.04 is probably affected by both an intrinsically higher ratio from closer to \ori\ and the wavelength dependent scattering by the PDR particles.  

The importance of the very different resolutions of the optical and radio data is probably more important here than it was for the comparison of full slit length samples. However, the convolution of the image by only the east-west component of the 327.5 MHz FWHM should have been adequate since the nebula is very similar along a north-south line in this region. 

\subsection{What these results tell us about the North-South Rim}

The North-South Rim feature delineates the east boundary of the EON. This feature is the sharpest boundary of the EON as images of the full EON show that the other portions of the boundary are either ill-defined or highly structured.
Examination of a combined infrared (3.6 $\mu$, and 8.0 $\mu$ and optical (0.43 $\mu$--0.9 $\mu$) (T. Megeath et~al. Spitzer Space Telescope release ssc2006-21a) shows that the North-South Rim is independent of the component of the Veil that forms the Dark Bay. Moreover, one can see the edge that forms the North-South Rim continues to the northeast and north of the Trapezium, but is interrupted by the Dark Bay, thus placing the North-South Rim between the Veil and the bright concave center of the Huygens region. If this is correct, then there would be expected to be some locally emitted emission lines because the feature would then be directly illuminated by LyC radiation from \ori\ and we know from the discussion in \S\ 3.6 that most of its 
emission line radiation does not originate locally. This fact would place the North-South Rim at a considerable distance from \ori, yet closer to \ori\ than the Veil.  This model is similar to a more schematic  construction for a more northerly portion of the North-South Rim prepared from quantitative emission-line and continuum imaging of the inner EON \citep{dop75} and consideration of there being two velocity
components of the [N~II] 6583 \AA\ line in this region\citep{deh73}. 

A better idea of the nature and location of the North-South Rim could be obtained from a series of narrow-band emission line images (it would expected that the locally emitted radiation will be displaced from the emission lines resulting from scattering) and/or high resolution spectroscopy (because of the expected velocity difference of the gas in the Huygens and North-South Rim regions). Explaining the existence of the North-South Rim is an open problem, the solution probably being related to those forces shaping the greater EON.

\subsection {Implications of the Important Role of Scattered Light}

This study has established that the Huygens region scattering of radiation is a general phenomenon. Moreover,  as the distance from \ori\ increases, scattered light becomes progressively more important. At sufficiently large separations scattered light can even be the dominant source of the observed radiation. This has grave implications for the use of emission lines to analyze the local physical conditions. It has already been argued that differences in the conditions for back-scattering \citep{od01}  (which will have a wavelength dependence and be dependent on the relative location of the emitting gas and the scattering dust) can modify the low velocity resolution emission line ratios and thus distort the derived physical conditions. The increasing importance of scattered light in the outer parts of the nebula means that derivation of conditions there are even more difficult. For example, an observed [S~II] doublet ratio would be a mix of local and scattered light and would produce only an upper limit to the local density.  The situation becomes even more difficult when the analysis must consider modification of the flux ratios due to the wavelength dependence of the scattered light, which means that derivation of electron temperatures from emission lines widely separated in wavelength proves to be very uncertain.

The problems illuminated here probably also have application in the analysis of emission lines from other H~II regions. The observational selection effect is for optical observations to identify objects with geometries similar to the Orion Nebula. If the H~II regions form deep within the parent molecular clouds or on sides not facing us, they will not be seen optically. Massive stars will probably not be formed in regions of low density at the edge of the molecular clouds. These biases favor our finding Orion-like H~II regions, neither deep within the molecular clouds and not so far out that there is still some overlying material.  If  the brightest parts of a nebula are spatially resolved, then the only concern is back-scattering. Fortunately, this can be done for many Galactic H~II regions, but when dealing with extra-galactic H~II regions observations are of integrated original and modified scattered emission, thus leaving moot the derivation of physical conditions and abundances.

\subsection{Comparison with the Results of Subrahmanyan, et~al. (2001)}

Our study has confirmed the conclusion of \citet{sgm} that the eastern boundary of the EON is dominated by scattered light, although they were not able to explicitly tie this to scattering of Huygens region radiation. Their conclusion that radiation from the entire outer boundary of the EON is dominated by scattered light is probably correct.  However, our conclusion that observed radiation from the central parts of the EON is a combination of local emission and scattered light from the Huygens region means that it is impossible with low velocity resolution to obtain an idea of the local densities and thickness of the EON emitting layer from optical data, while combinations of optically derived densities with radio EM values produce only limits. For example, the 327.5 MHz EM for the S840 sample is 2,180 \emunits\ and the [S~II] derived density is 110 \cmq, which must be an upper limit to the density since part of the [S~II] emission will be scattered light from the much higher density Huygens region. The lower limit to the thickness of the S840 sample calculated assuming a constant density  would be 0.2 pc, similar to that found for the Huygens region.  

\section{Conclusions}

We can draw several important conclusions from this study.

1. There is a steady progression of characteristics of the EON with increasing distance from \ori, including changes in the F(\Halpha)/F(\Hbeta) ratio, the Equivalent Width of the continuum at \Hbeta, the density derived from the [S~II] doublet, and the ratio of Emission Measures derived from the optical \Hbeta\ line and the 327.5 MHz radio continuum.

2. The relative strength of scattered light from stellar continuum becomes more important with distance from the Trapezium.

3. Scattered light from the bright Huygens region is an important contributor to emission line radiation from the EON, thus confusing attempts to determine the physical conditions there. 

4.The major role played by scattered emission line radiation in the EON strengthens earlier arguments that scattered light in Huygens region low resolution spectroscopy imposes a fundamental limit on our ability to determine the physical conditions and abundances in that region.

5. The North-South Rim that forms the eastern boundary of the EON is primarily seen through scattered light originating from the Huygens region and possibly nearby portions of the EON.

Many of our conclusions were anticipated in a discussion of the large scale structure of the Orion Nebula and its surrounding region \citep{pei82} utilizing the data available then. Since that time the importance of scattering from the background PDR rather than enmeshed dust has come to be appreciated. The imaging study of \citet{dop75} came to very similar conclusions about the importance of scattered light and the North-South Rim feature, although in both cases only addressing the most northerly portion of the EON.

\acknowledgments
The useful comments of Manuel Peimbert  and Gary J. Ferland on a draft of this paper are gratefully acknowledged. We are also grateful to Jose Velasquez  of the CTIO staff for assistance in obtaining the optical spectra. The National Radio Astronomy Observatory  is a facility of the
National Science Foundation operated under cooperative agreement
with Associated Universities, Inc.
Partial financial support for CRO's work on this project was provided by STScI grant GO 10967 (CO, Principal Investigator) and Spitzer Space Telescope grant GO50082 (Robert H. Rubin, Principal Investigator) .

{\it Facilities:} \facility{CTIO(1.5 m)}, \facility{VLA}.

\begin{deluxetable}{cccccc}
\tabletypesize{\scriptsize}
\tablecaption{Results from Spectroscopy}
\tablewidth{0pt}
\tablehead{
\colhead{Sample} & \colhead{Distance-\ori~(\arcsec)} & \colhead{S$^{1}_{H\beta}$} & \colhead{F(6716 \AA)/F(6731 \AA)} & \colhead{Equivalent Width (\AA)} & \colhead{F(\Halpha)/F(\Hbeta)}}
\startdata
S240    & 240  & 4.69x10$^{8}$ & 1.04 & 330 & 3.16\\
S360    & 360  & 1.70x10$^{8}$ & 1.16 & 240 & 2.86\\
S420    & 422  & 1.89x10$^{8}$ & 1.20 & 270 & 2.70\\
S480    & 482  & 1.37x10$^{8}$ &  1.22 & 350 & 2.75\\
JW75   & 528  & 1.92x10$^{8}$ &  1.27 & 460 & 2.85\\
JW887 & 563  & 1.30x10$^{8}$ & 1.24 & 250 & 2.86\\
S600    &  601 & 1.22x10$^{8} $& 1.29 & 350 & 2.72\\
S720    & 721  & 7.43x10$^{7}$&  1.33 & 260 & 2.77\\
P1605  & 802  & 6.51x10$^{7}$ & 1.32 & --- & 2.74\\
S840    & 841  & 5.12x10$^{7}$ & 1.32 & 210 & 2.64\\
P1353  &1049 & 7.44x10$^{7}$ &1.31 & 400 & 3.09\\
\enddata
\tablecomments{$^{1}$ S$_{H\beta}$ is in units of photons \cmsq\ s$^{-1}$ steradian$^{-1}$}
\end{deluxetable}

\begin{deluxetable}{ccc}
\tabletypesize{\scriptsize}
\tablecaption{Parameters Derived from Spectroscopy}
\tablewidth{0pt}
\tablehead{
\colhead{Sample} & \colhead{N$^{1}_{e}$([S~II])} & \colhead{Emission Measure$^{2}$}}
\startdata
S240   &    500 & 52600\\
S360   &   290 & 19100\\
S420   &   240 & 21200\\
S480   &   210 & 15400\\
JW75   & 150  & 21600\\
JW887 & 190 & 14500\\
S600   & 140  & 13700\\
S720   &   90  &  8350\\
P1605 & 110 & 7300\\
S840   &  110 & 5750\\
P1353 & 110 & 8350\\
\enddata
\tablecomments{$^{1}$Density derived from the [S~II] doublet ratio (\cmq), $^{2}$Emission Measure units are \emunits}
\end{deluxetable}

\begin{deluxetable}{ccc}
\tabletypesize{\scriptsize}
\tablecaption{Parameters Derived from the 327.5 MHz Image}
\tablewidth{0pt}
\tablehead{
\colhead{Sample} & \colhead{Emission Measure$^{1}$} & \colhead{EM(\Hbeta)/EM(327.5~MHz)}}
\startdata
S240   &  4.36x10$^{4}$ & 1.21 \\
S360   &  1.80x10$^{4}$ & 1.06 \\
S420   &    9.92x10$^{3}$ &  2.14\\
S480   &    9.24x10$^{3}$ & 1.67\\
JW75   &  2.33x10$^{4}$ & 0.93\\
JW887 &  8.74x10$^{3}$ &  1.66 \\
S600   &   8.87x10$^{3}$ &  1.55 \\
S720   &   3.84x10$^{3}$ &   2.18\\
P1605 &   3.35x10$^{3}$ &   2.18\\
S840   &    2.17x10$^{3}$  & 2.65\\
P1353 &    6.17x10$^{3}$  & 1.35 \\
\enddata
\tablecomments{$^{1}$Emission Measure units are \emunits}
\end{deluxetable}

\begin{figure}
\epsscale{1.0}
\plotone{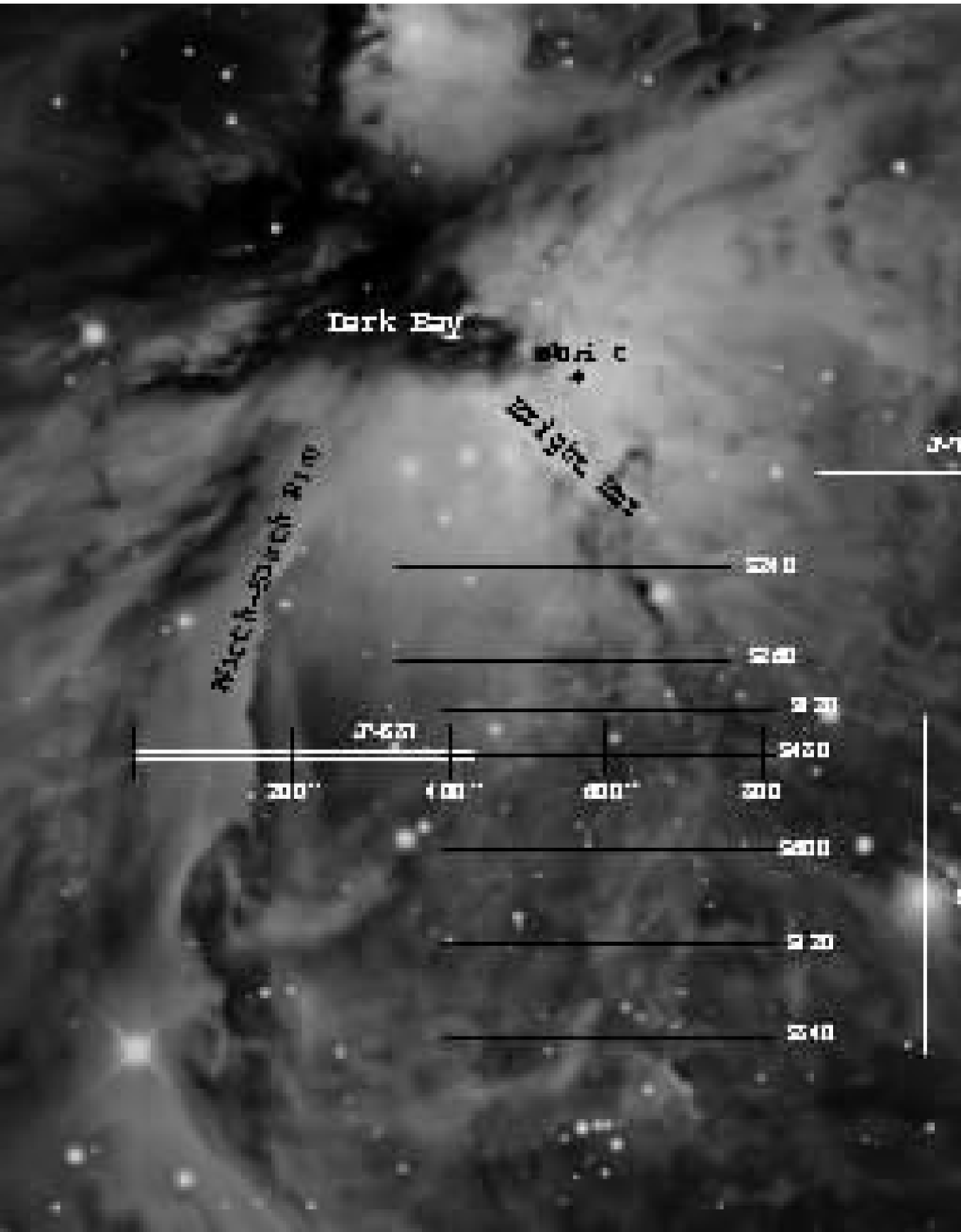}
\caption{
This 2182\arcsec\ 1563\arcsec\  groundbased telescope image using broad bandpass filters approximating the UBV system has superimposed the positions of the slits used for our spectroscopic study.
The slits marked JW 887 and S480 were used to derive the profile shown in Figure 3 and the distances along that profile are labeled in this figure. Image used with the permission of Robert Gendler (http://www.robgendlerastropics.com). \label{fig:slits}}
\end{figure}

\begin{figure}
\epsscale{1.0}
\plotone{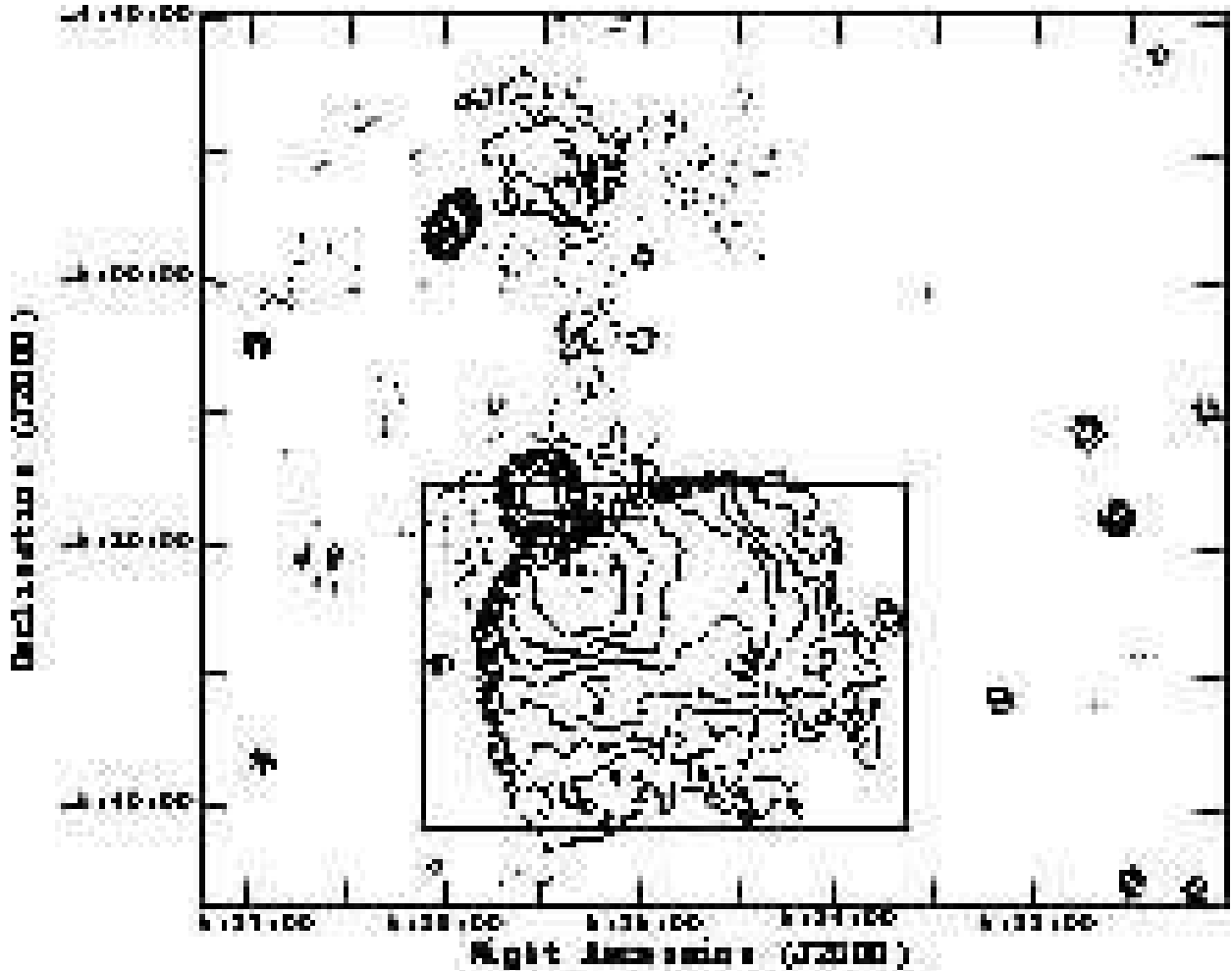}
\caption{
This figure shows the VLA image at 327.5~MHz with a resolution  $78.6\arcsec \times 65.0\arcsec$ at position angle $25^{\arcdeg}$ \citep{sgm}. The enclosed rectangle designates the area covered in the optical image in Figure 1 and the position of \ori\ has been indicated. The rms noise is 7.6 mJy~beam$^{-1}$. The contour units are -32, -16, 16, 32, 64, 128, 256, 512, 1024, and 2048 mJy~beam$^{-1}$. The peak in the VLA image at 327.5~MHz is 3.23 Jy~beam$^{-1}$. NGC 1973-75-77 is located about one half degree north of the main complex of NGC 1976 and NGC 1982 . 
\label{fig:radio}}
\end{figure}

\begin{figure}
\epsscale{1.0}
\plotone{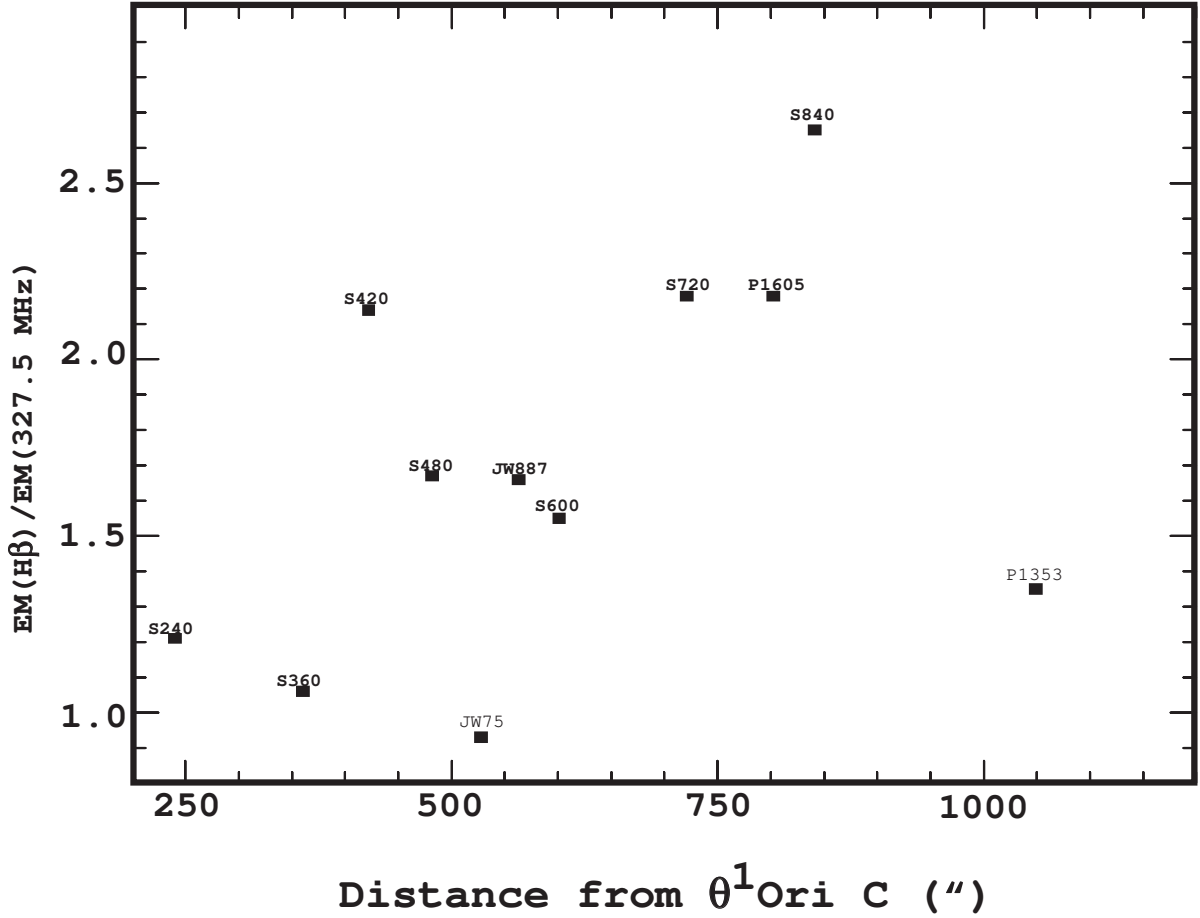}
\caption{
The ratio of the emission measure derived from the optical (\Hbeta) and 327.5 MHz observations
and the distance from \ori are shown. The ratios shown are derived from along all of each slit sample and the distance shown is for the center of the slit samples. The bold face names are for samples along a south pattern and the light face names are for samples extending to the west.
 \label{fig:distanceratio}}
\end{figure}

\begin{figure}
\epsscale{1.0}
\plotone{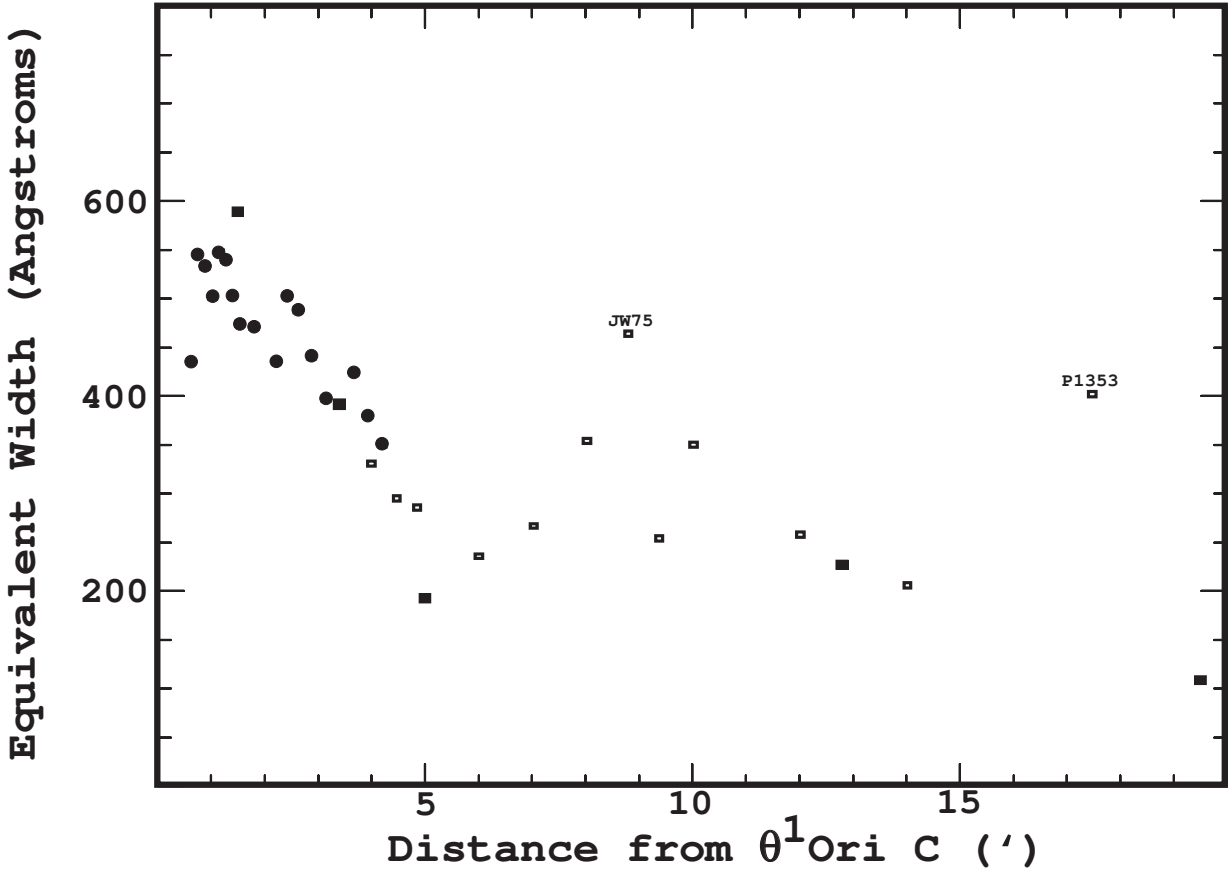}
\caption{The observed Equivalent Width at \Hbeta\ is plotted as a function of distance from \ori\ from three sets of observations. Filled circles are from \citet{b91}, filled rectangles from \citet{oh65}, and open rectangles are from this study. 
\label{fig:EW}}
\end{figure}

\begin{figure}
\epsscale{1.0}
\plotone{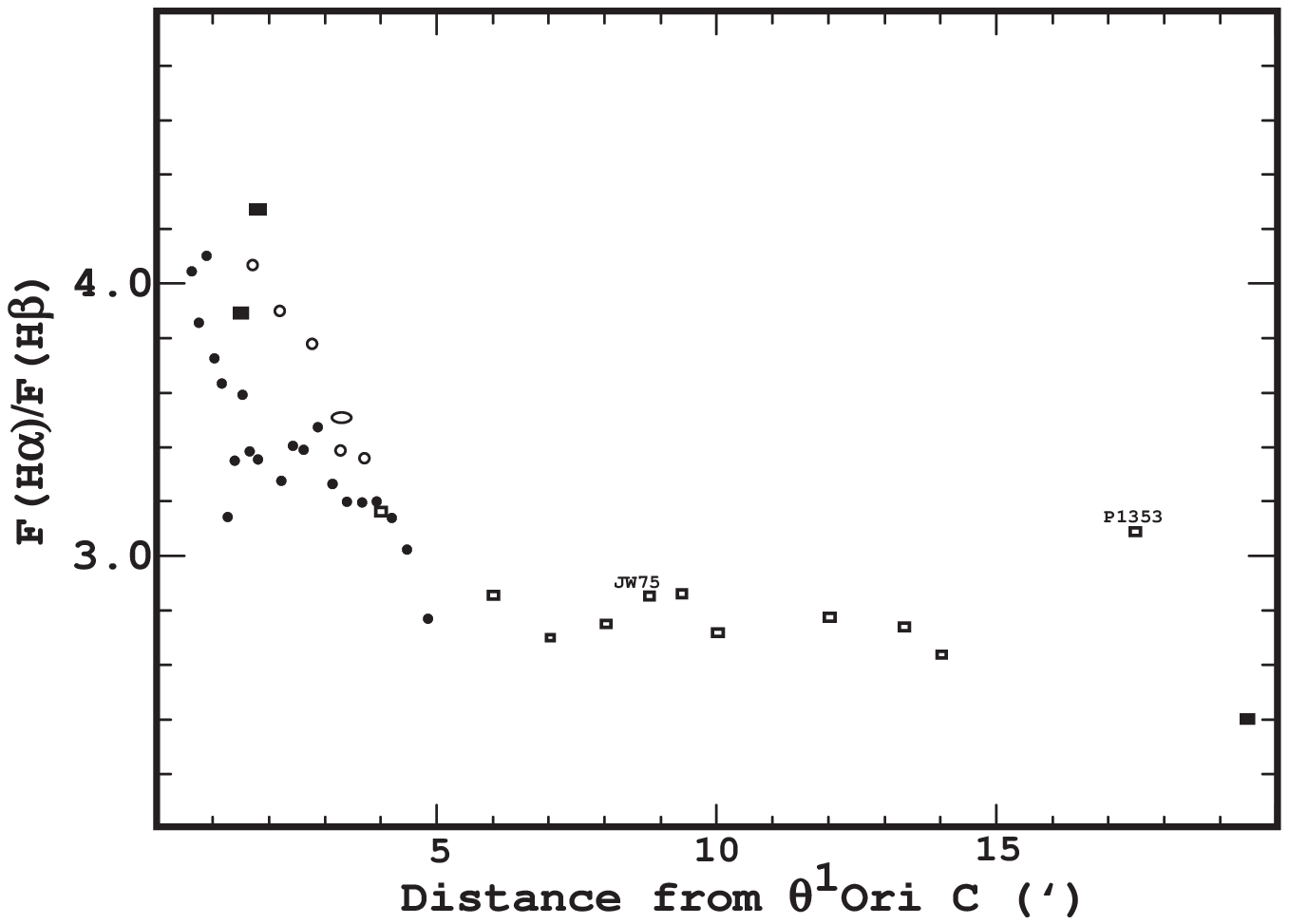}
\caption{
The observed flux ratio of the first two Balmer lines is plotted against distance from \ori.
Filled rectangles are from \citet{oh65}, open rectangles from this study, filled circles from \citet{b91}, the open ellipse from \citet{jps73},  and the five open circles are the southern samples of \citet{pp77}.
\label{fig:Balmer}}
\end{figure}

\begin{figure}
\epsscale{1.0}
\plotone{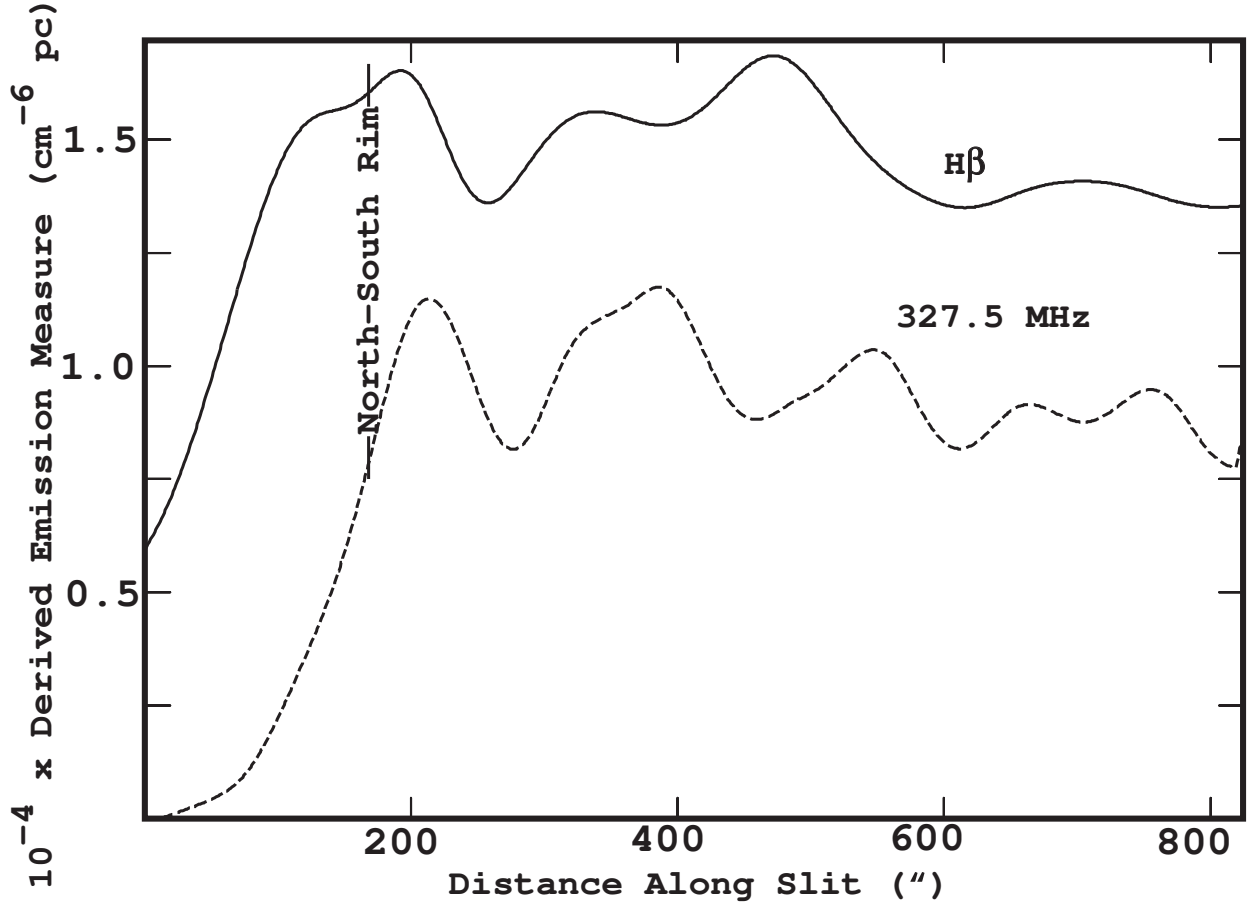}
\caption{
The emission measure derived from the optical (\Hbeta) and 327.5 MHz observations are presented as a function of distance from the east end of slit JW 887. An electron temperature of 9000 K was assumed and the \Hbeta\ profile has been gaussian convolved to a FWHM of 67\arcsec. The feature designated as the North-South Rim in Figure 1 occurs at a distance of 168\arcsec, as shown.\label{fig:both}}
\end{figure}

\begin{figure}
\epsscale{1.0}
\plotone{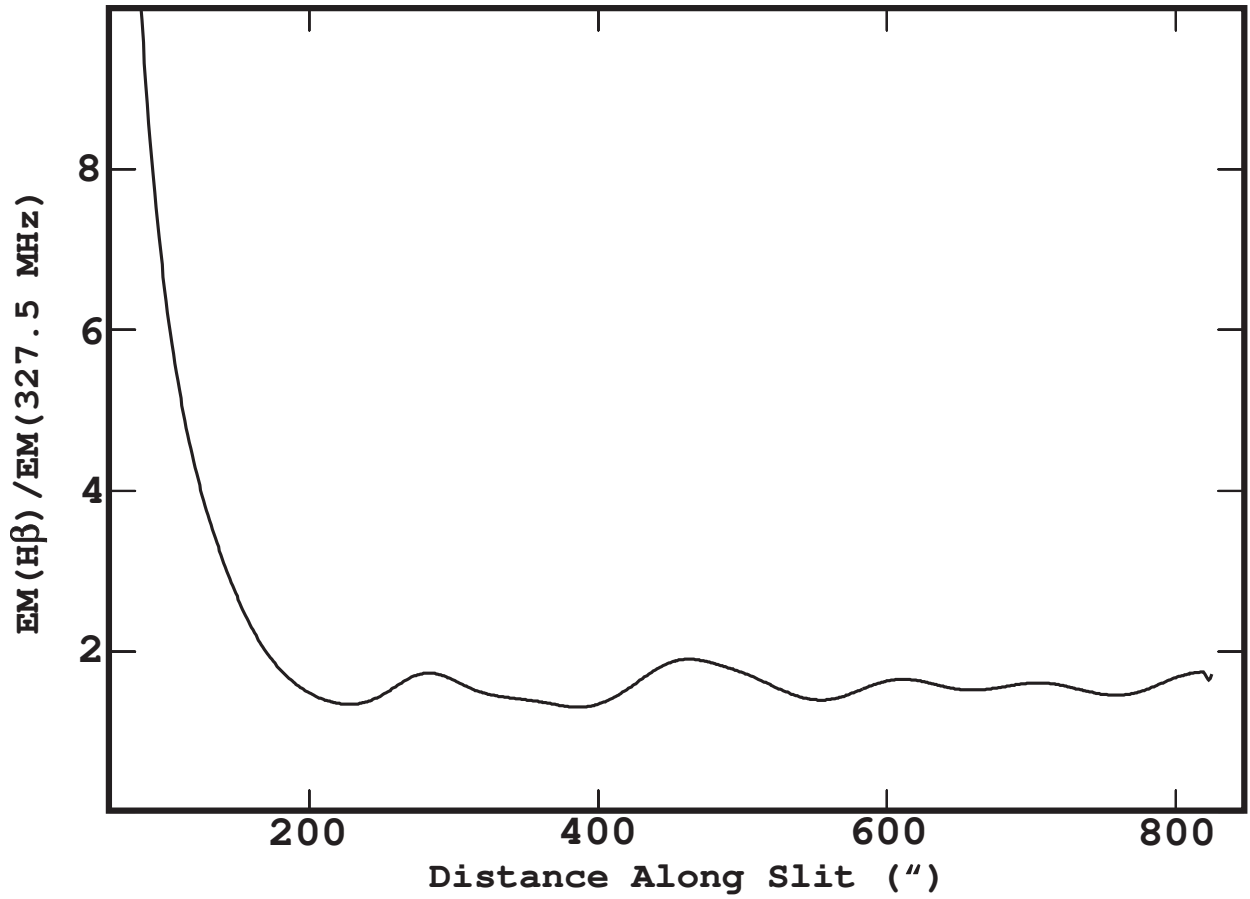}
\caption{
The ratio of the emission measure derived from the optical (\Hbeta) and 327.5 MHz observations are presented as a function of distance from the east end of slit JW 887. T
 \label{fig:ratio}}
\end{figure}

\end{document}